\documentclass{IEEEcsmag}

\usepackage[hyphens]{url}
\usepackage[colorlinks,urlcolor=blue,linkcolor=blue,citecolor=blue]{hyperref}
\hypersetup{breaklinks=true}
\usepackage{graphicx}
\usepackage{upmath}

\jvol{XX}
\jnum{XX}
\paper{X}
\jmonth{March/April}
\jname{Computing in Science and Engineering}
\pubyear{2023}
\begin{document}

\sptitle{CiSE: Lorena Barba} 
\editor{Editor: Jeffrey C. Carver, carver@cs.ua.edu}
 
\title{Research Software Science: Expanding the Impact of Research Software Engineering}

\author{Michael~A. Heroux}
\affil{Saint John's University, Sandia National Laboratories}

\markboth{CiSE Special Issue on the Future of Research Software Engineers in the US}{}

\begin{abstract}
Software plays a central role in scientific discovery. Improving how we develop and use software for research can have both broad and deep impacts on a spectrum of challenges and opportunities society faces today. The emergence of Research Software Engineer (RSE) as a role correlates with the growing complexity of scientific challenges and diversity of software team skills.  In this paper, we describe research software science (RSS), an idea related to RSE, and particularly suited to research software teams.  RSS promotes the use of scientific methodologies to explore and establish broadly applicable knowledge. Using RSS, we can pursue sustainable, repeatable, and reproducible software improvements that positively impact research software toward improved scientific discovery.
\end{abstract}

\maketitle
  
\chapterinitial{Research Software Engineer (RSE)} has emerged as an identity for many members of the research software community~\cite{RSE-Society,US-RSE}.  For many years, RSE functions were needed on scientific teams but only recently has there been the growth in awareness of the importance of these skills and the people who possess them, leading to a long-overdue and explicit recognition that RSE staff require stable career paths and communities of their own, beyond solely being part of a particular scientific team.

RSE job functions rely on the premise that there are better and worse ways to produce software for use in scientific research.  Often, RSEs read the existing software engineering literature to keep informed about, adapt, and adopt evolving practices and tools.  They also carry techniques and experience from project to project.  These translational strategies are intrinsically valuable and are a part of the fundamental value proposition that RSEs bring to the scientific community.

\begin{figure*}[h]
	
	\centering
	\includegraphics[scale=0.2]{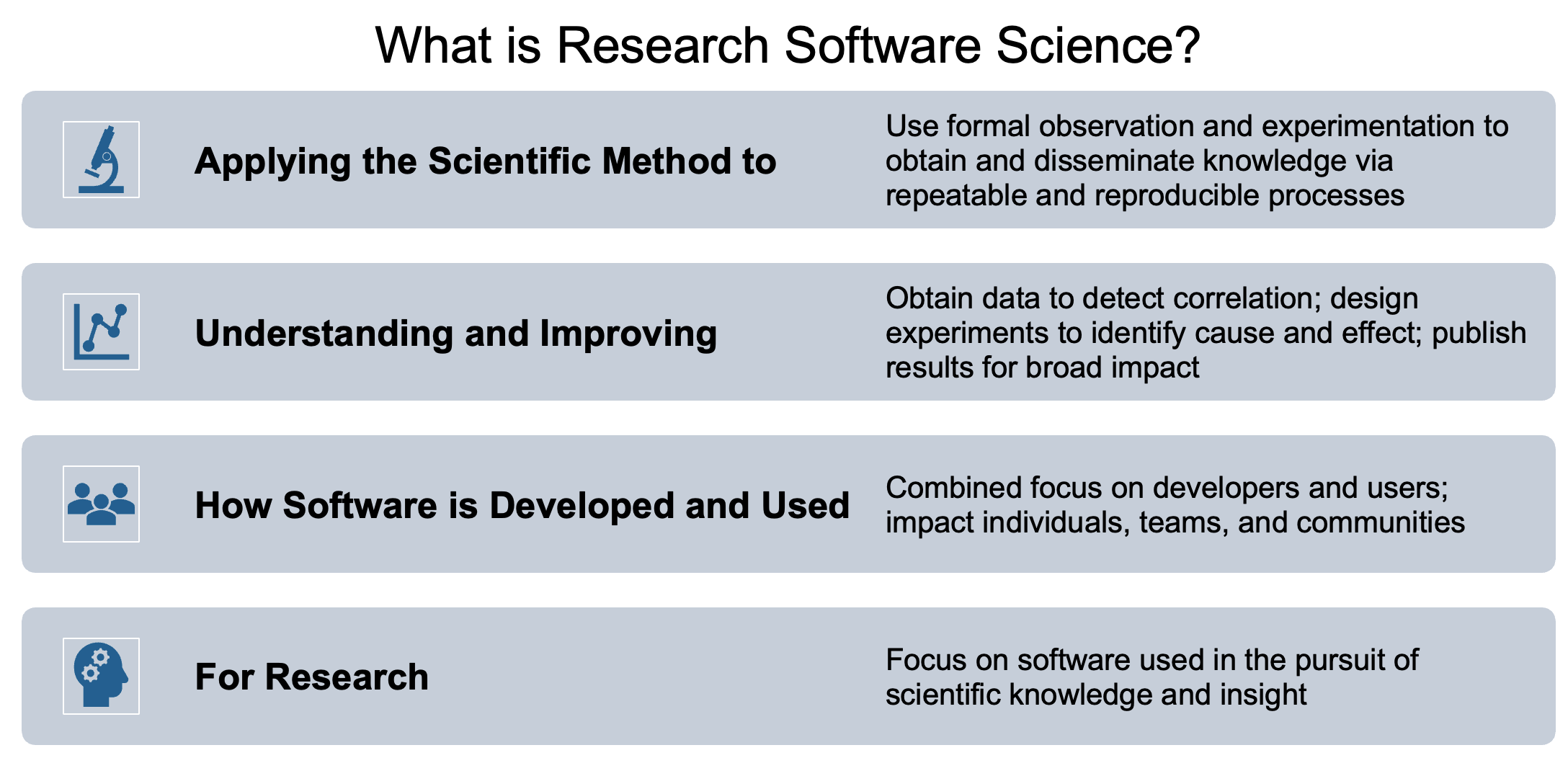}
	\caption{\footnotesize Research software science (RSS) is proposed as a complementary approach to RSE efforts for improving how software is developed and used for research.  RSS leverages the inherent appreciation for scientific methodologies present in the research community, providing another source of information for RSE practitioners to leverage.}
	\label{ResearchSoftwareScienceDefinition}
\end{figure*}

In this paper we present the concept of \textit{research software science} (RSS): applying the scientific method to understanding and improving how software is developed and used for research (Figure~\ref{ResearchSoftwareScienceDefinition}).  We argue that RSS is an important complement to RSE efforts, providing another avenue to enhancing the impact of RSE efforts beyond translational activities.  Furthermore, RSS leverages the innate scientific sensibilities of the research communities to which RSE members belong.  In other words, while it is appropriate to apply a scientific approach to understanding and improving how any kind of software is developed and used, it is particularly apt to use this approach for research software. Finally, by emphasizing a scientific approach to improving software activities, research funding organizations can more readily justify investments in advancing how their sponsored software projects improve the practice of developing and using software for research, as can be seen by the recent US Department of Energy sponsorship of the Workshop on the Science of Scientific-Software Development and Use~\cite{Bernholdt:2022:SSS} in December 2021.

\section{Background}
Development and use of software are fundamental to numerous areas of scientific research. Many scientists write, modify, and use software to gain insight and prove scientific results. At the same time, formal software engineering techniques and knowledge that are widely adopted in other software development domains are not as commonly used in research software projects. In our experience, research software development approaches are more informal, particularly in the upstream activities of requirements, analysis, and design.

One challenge to investing in improved software practices and processes for science is the perception by some that a focus on improving software skills falls under the category of engineering, not science. There is a perception that software engineering—refining repeated practices for more efficiency—is not something that a scientific research funding portfolio should support as a fundamental element.  

From this perspective, the best that research software teams can do is spend modest amounts of time learning practices from the mainstream software engineering literature and use them unchanged or with modest adaptation in their own software development. While the RSE movement is improving the situation for larger scientific team, in our experience, the current status results in only moderate success—and sometimes even failure. Best practices distilled from larger, more mainstream software domains may be ill-suited for research software teams. 

The growth of the RSE community is a strong sign that these perceptions are changing, and software engineering is increasingly recognized as an essential skill set on a diverse scientific software team.  Even so, there is another opportunity to improve our ability to develop software and improve its usability.

In this article, we propose that the scientific method—which is central to scientific efforts using research software—can be used to study and improve the development and use of research software. Looking at the development and use of software for research through a scientific lens enables us to apply a scientific approach to understand and improve software as a tool for research. In other words, research software development and use are the subjects of our scientific study.

\section{Science applied to research software}
As stated above, the primary objective of research software science as we are proposing it is to apply the scientific method to understanding the development and use of research software. This pursuit has strong technical, social, and cognitive components, describe below.

\paragraph{Technical component:}
The purpose of research software is modeling and simulation of scientific theories; the gathering, analysis, and understanding of scientific data; and related pursuits. These activities typically require years of education in a scientific domain and ongoing community engagement to contribute to and keep abreast of new discoveries and approaches. For example, to develop and use computational fluid dynamics (CFD) research software, one must complete years of study in mathematics, physics, and engineering and then continue study of CFD and related fields even when developing and using software.

\paragraph{Social component:}
Development and use of scientific software are typically a team effort, increasingly involving more people and more diverse roles. Team interactions, workflows, and tools play a large part in the effectiveness of a research software team. While many people have realized the importance of the technical component of research software development and use, fewer people have focused on the social elements, and even fewer have applied a scientific approach to studying and improving research software team interactions.

\paragraph{Cognitive component:}
Improving our approaches to developing and using research software typically requires learning. In our experience, scientists tend to enjoy solving problems. Framing change as a problem or puzzle to be solved can be effective in engaging scientists. Posing improvement goals in a descriptive way, more than a prescriptive, enables scientists to be part of the creative process. More generally, leveraging knowledge from cognitive sciences improves our ability to understand how developers and users of research software approach their work and interact with each other.

\section{Social and cognitive sciences focus}
Applying the scientific method to research software teams necessarily involves the social and cognitive sciences. Observations, interviews, data mining, and similar techniques provide the raw materials for analyzing and gaining understanding of important correlations—and ultimately, one hopes, identifying cause and effect—between behaviors, situations, and outcomes.

While the software engineering literature addresses the social and cognitive elements of software development and use, research software teams take on considerable risks by adopting published team practices without scrutiny or adaptation. In our experience, many published team practices are not sufficiently informed by the dynamics and requirements of research software development and use. To better understand when and how existing software methodologies are appropriate and to develop new approaches for research software, we need the skills and tools of social and cognitive sciences applied to research software teams and individuals.

\section{RSS is more than just an extension of RSE}
As mentioned, the software engineering literature provides ample material dedicated to the social and cognitive elements of software teams. Some of this literature is scientific in nature, but much of it is anecdotal, based on years of experience from seasoned software professionals. These writers produce generalized recommendations from specific experiences, often with benefit to other developers and users.
\begin{figure*}[h]
	
	\centering
	\includegraphics[scale=0.2]{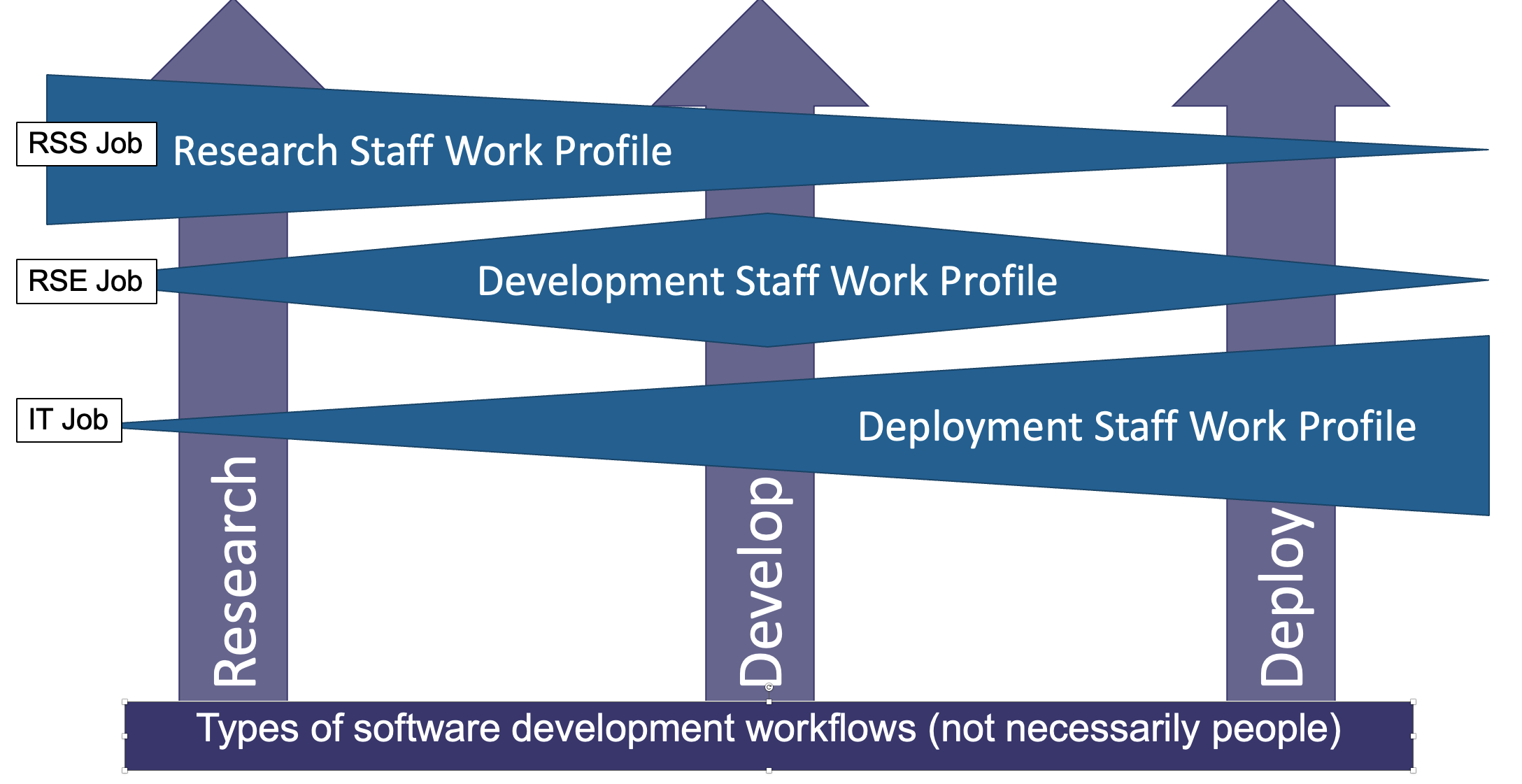}
	\caption{\footnotesize One model for integrating research software science staff into an existing research software engineering teams is to consider a research-develop-deploy pipeline.  Opportunities for improving development and use of research software approaches come from understanding the challenges faced in deploying existing software products.  These challenges become the research problems for study and later become new approaches to develop and deploy.}
	\label{ResearchDevelopDeployWorkflows}
\end{figure*}

It is reasonable to argue that RSS is a modest extension of RSE.  There is some truth to this, especially to the extent RSE team members experiment with new mental models, tools, and processes.  However, in our experience, this experimentation is seldom a formal, repeatable, or reproducible process designed to generate sharable knowledge.  Instead, we believe it is fruitful to consider RSS as new identifiable element within an existing RSE organization.  Figure~\ref{ResearchDevelopDeployWorkflows} shows a notional research-develop-deploy pipeline where the challenges of supporting research software within the deploy phase inform the  questions in the research phase that are then used to inform the develop phase and lead to new deployable capabilities.

Fred Brooks is quoted as saying, ``A scientist builds in order to learn; an engineer learns in order to build.''  Engineers want an improved tool or process.  They identify a few possibilities, test, select the best, and move on.  There is only incidental team memory and little focus on dissemination. Scientists want to understand underlying principles, correlation, cause-and-effect.  They design studies, capture data, analyze results, and publish.

Clearly, the software engineering community performs research, but this research is not always directly applicable to research software.  Furthermore, especially in the research community, we should call this kind of work what it is: science.
\begin{figure*}[h]
	
	\centering
	\includegraphics[scale=0.25]{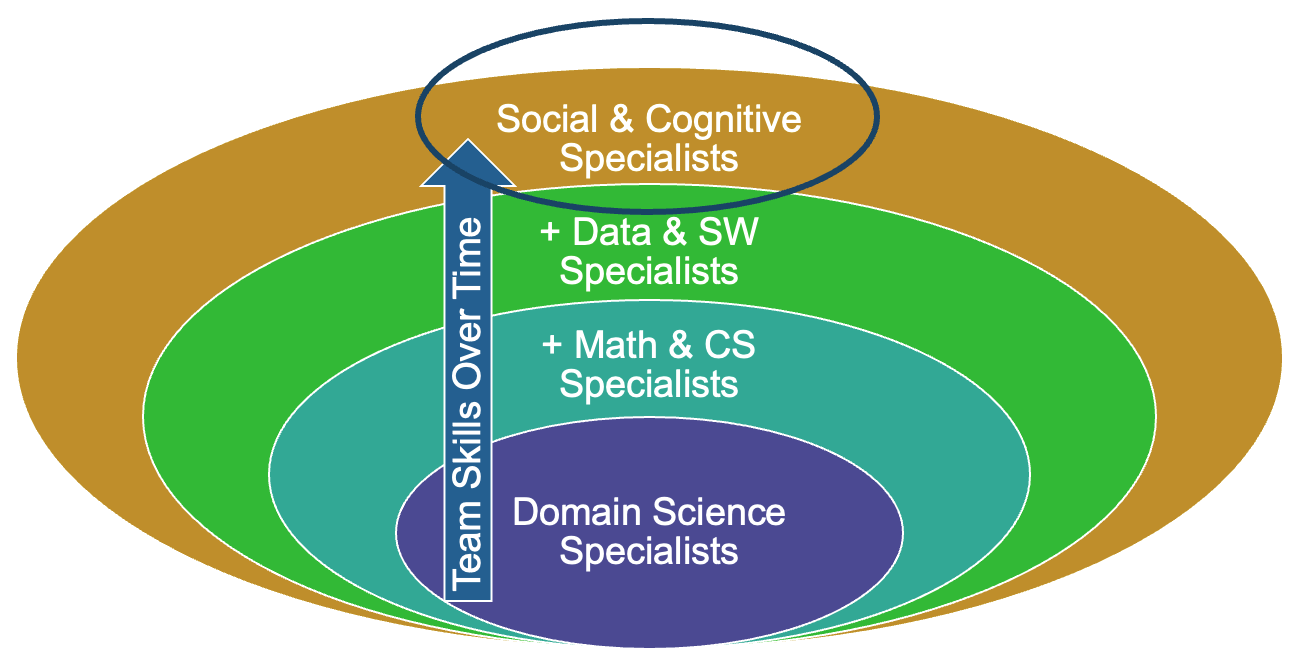}
	\caption{\footnotesize Early use of software for scientific discovery was generally small scale, developed by a team of domain scientists.  As efforts proceeded, the need for new mathematics and more efficient computer science algorithms and data structures emerged. As software products grew in size, complexity and coupling, the need for data management and software engineering expertise emerged.  Presently, the growing complexity of scientific software projects increases the value of investing in cognitive and social sciences.}
	\label{TeamSkillSetsOverTime}
\end{figure*}

\section{Why now: Multidisciplinary direction of science}
Many important efforts in science require strong multidisciplinary teams. We see that research software is increasingly incorporating multiscale and multiphysics features, equation-driven and data driven approaches, involving modeling, simulation, and data analysis. Adding a scientific approach to understanding the development and use of research software establishes one more dimension in the pursuit of better scientific approaches and is especially appropriate given the growing diversity of scientific teams and the need to understand and optimize team interactions and output.  Historically, we can see the expansion of skillsets needed by leading research software communities, as shown in Figure~\ref{TeamSkillSetsOverTime}.

\section{Trends in scientific software that increase the value of RSS}
The technology community is seeing a growing importance for considering human factors in product development [cite Tett].  In addition, software design and development platforms are becoming increasingly sophisticated, reducing the cost of creating products.  For example, artificial intelligence (AI)-based tools that assist in generating source code, testing infrastructure, and more, are available to assist in producing software.  

With the increased emphasis on improving the usability of software and the reduced cost of producing it, skills in eliciting and analyzing requirements, and user-centered design become relatively more important than development skills. More time will be spent on the upstream and higher-level activities of what the product should be than on making the product.  Because of these trends, we have an opportunity to place more emphasis on purpose and design, providing software systems adapted to fit scientists, broadening usability, accessibility, and impact.

As part of the trends we observe, software engineering focuses such as user experience (UX) can make more sense in the research software development process to address the growing size and complexity of scientific teams and environments.  Research software science can play a large role in guiding these UX activities, providing a scientific foundation for long-term and sustainable impact. One notable UX opportunity for scientific software is that on many teams the users are also the primary developers, an atypical situation for most UX methodologies.

\section{CONCLUSION}

Development and use of research software are rich and dynamic pursuits, worthy of scientific study in their own right. Viewing the improvement challenges as scientific problems opens the door to applying scientific skills to assist in making our software development and use even more effective.

Research software communities of all kinds expect rigorous use of the scientific method as we produce and gain scientific insight from the development and use of software in our domains of study; fluid dynamics research, for example, has benefited tremendously from computation.
We should expect the same rigor as we try to understand how to better develop and use the software tools that enable our research. Research software science—applying science to the study of how we develop and use research software—should qualitatively improve our software capabilities. Forming and promoting the role of a research software scientist should lead to the same qualitative improvement in outcomes we have seen in any other domain where the scientific method has taken hold. Because software is such an important element of research, we expect that focusing on research software science should dramatically improve science overall.

\section{ACKNOWLEDGMENT}

The original idea of Research Software Science presented in this article was first delivered as a white paper at the 2019 Collegeville Workshop on Sustainable Scientific Software (CW3S19)~\cite{Collegeville2019}, then revised and extended for a blog post~\cite{Heroux:2019:RSS}, cross posted on the \href{https://bssw.io}{BSSw} and \href{https://urssi.us}{URSSI} websites. We thank St. John's University for the on-site and virtual resources used to conduct the Collegeville Workshops on Scientific Software~\cite{CollegevilleSeries} that were the genesis of the concept of Research Software Science.


\begin{IEEEbiography}{Michael A. Heroux}{\,} is a Scientist in Residence at St. John's University, MN, a Senior Scientist at Sandia National Laboratories, and Director of Software Technology for the US Department of Energy (DOE) Exascale Computing Project (ECP). He is the founder of the Trilinos scientific libraries, Kokkos performance portability libraries, Mantevo miniapps and HPCG Benchmark projects, and is presently leading the Extreme-scale Scientific Software Stack (E4S) project in DOE, a curated collection of HPC software targeting leadership platforms. Mike is a Fellow of the Society for Industrial and Applied Mathematics (SIAM), a Distinguished Member of the Association for Computing Machinery (ACM), and a Senior Member of IEEE. Contact him at mheroux@csbsju.edu.
\end{IEEEbiography}

\end{document}